\documentclass[a4paper]{article}

\usepackage{INTERSPEECH2020}
\usepackage{acronym}
\usepackage[table,x11names]{xcolor}
\usepackage{floatrow}
\usepackage{hyperref}
\hypersetup{
    colorlinks=true,
    linkcolor=blue,
    urlcolor=magenta,
}

\newcommand{\etal}{\emph{et al.} }

\acrodef{SNR}{signal-to-noise ratio}
\acrodef{STRF}{spectro-temporal receptive field}
\acrodef{GRU}{gated recurrent unit}
\acrodef{CQT}{constant-Q transform}
\acrodef{CNN}{convolutional neural network}
\acrodef{MLP}{multi-layer perceptron}
\acrodef{VAD}{voice activity detection}

\title{Learnable Spectro-temporal Receptive Fields for\\Robust Voice Type Discrimination}
\name{Tyler Vuong,$^{1}$
      Yangyang Xia,$^{1}$
      Richard M. Stern$^{1,2}$}
      
\address{$^1$ Department of Electrical and Computer Engineering, Carnegie Mellon University, USA\\
         $^2$ Language Technologies Institute, Carnegie Mellon University, USA
}
\email{tvuong@andrew.cmu.edu, raymondxia@cmu.edu, rms@cs.cmu.edu}

\begin{document}

\maketitle
\begin{abstract}
Voice Type Discrimination (VTD) refers to discrimination between regions in a recording where speech was produced by speakers that are physically within proximity of the recording device (“Live Speech”) from speech and other types of audio that were played back such as traffic noise and television broadcasts (“Distractor Audio”). In this work, we propose a deep-learning-based VTD system that features an initial layer of learnable spectro-temporal receptive fields (STRFs).  Our approach is also shown to provide very strong performance on a similar spoofing detection task in the ASVspoof 2019 challenge. We evaluate our approach on a new standardized VTD database that was collected to support research in this area. In particular, we study the effect of using learnable STRFs compared to static STRFs or unconstrained kernels. We also  show that our system consistently improves a competitive baseline system across a wide range of signal-to-noise ratios on spoofing detection in the presence of  VTD distractor noise.
\end{abstract}

\section{Introduction}
The goal of Voice Type Discrimination (VTD) is to locate regions of ``live speech'' in an audio recording containing both ``live speech'' and  ``distractor audio.''  Live speech refers to speech segments spoken by speakers physically in the proximity of the recording device.  Distractor audio is any other audio source that is not live speech such as environmental noises, television, and radio.  This becomes challenging since distractor audio can contain natural human speech that is broadcast through a television or radio and often overlaps with the live speech in time. In addition, no assumption is made about the \ac{SNR} or the geometry of the room where the recording took place, making methods that use auxiliary hardware to detect specific live speech patterns \cite{shiota2016voice,sahidullah2017robust} inapplicable. VTD is important for practical applications such as wake-up word detection for voice assistants \cite{wu2018monophone}. Robust detection of voice under noise and music is also vital for defending against adversarial attack using nonspeech audio signals \cite{li2019adversarial}.

Carnegie Mellon University (CMU) was a participant in an assessment of the capabilities of VTD discrimination systems conducted in June and July of 2019 by the Johns Hopkins Applied Physics Laboratory (JHU/APL) using audio data that were recorded and annotated by SRI International (SRI)\cite{RicheyEtAl19} in four rooms of different size and shape.  We describe the latest CMU system in this paper.

The upper panel of Figure 1 depicts Room 3 used for the SRI data collection.  The lower panel of that figure is a color-coded timeline that describes the actual presence and absence of the live speech and the various distractors.  We note that live speech is present only a very small fraction of the total time.

\textbf{Related work.}  While this work is part of the first effort ever to develop systematic approaches to voice type discrimination, VTD itself has some resemblances to \ac{VAD} \cite{KidaEtAl06} and spoofing detection. Specifically, VTD can be thought of as a two-stage problem: (1) discrimination between speech and nonspeech, and (2)  the separation of live speech from machine speech. The latter is obviously the far more challenging problem. Recent advancement in automatic detection of spoofing attacks have focused on specific forms of countermeasures in synthetic, converted, and replayed speech \cite{Todisco2019}.  Related methods for each type of attack is therefore highly specialized. For example,  Replay Attack (RA) countermeasures typically rely on detecting distortions in the higher-frequency bands (\emph{e.g.} \cite{LiEtAl17}).  In VTD, however, we focus on more common distractor speech in TV and radio broadcasts. Despite being recorded in nature, broadcast speech contains a richer variety of speaking styles. In addition, the acoustical conditions for VTD  are often more adverse than for spoofing detection.

The concept of the \ac{STRF}  has been  applied  successfully in \ac{VAD} and representation learning of speech. The form of the STRFs is motivated by physiological structures  that are believed to exist in the central auditory system that respond to a range of patterns of temporal  modulation and spectral modulation \cite{WangShamma95}. Mesgarani \etal proposed a \ac{STRF}-based voice activity detector that is highly robust in the presence of excessive noise and reverberation \cite{mesgarani2006discrimination}. More recently, Gabor-based modulation filters were implemented as kernels in a \ac{CNN} to aid representation learning for robust speech recognition \cite{agrawal2019modulation}. In general, kernelizing \ac{CNN} layers as filters has also proven useful in speaker and phoneme recognition \cite{RavanelliBengio17,LoweimiEtAl19}. This work motivated us to build our VTD system around learnable \ac{STRF}s.


\begin{figure}[ht]
\centering
\includegraphics[width = 2.25in]{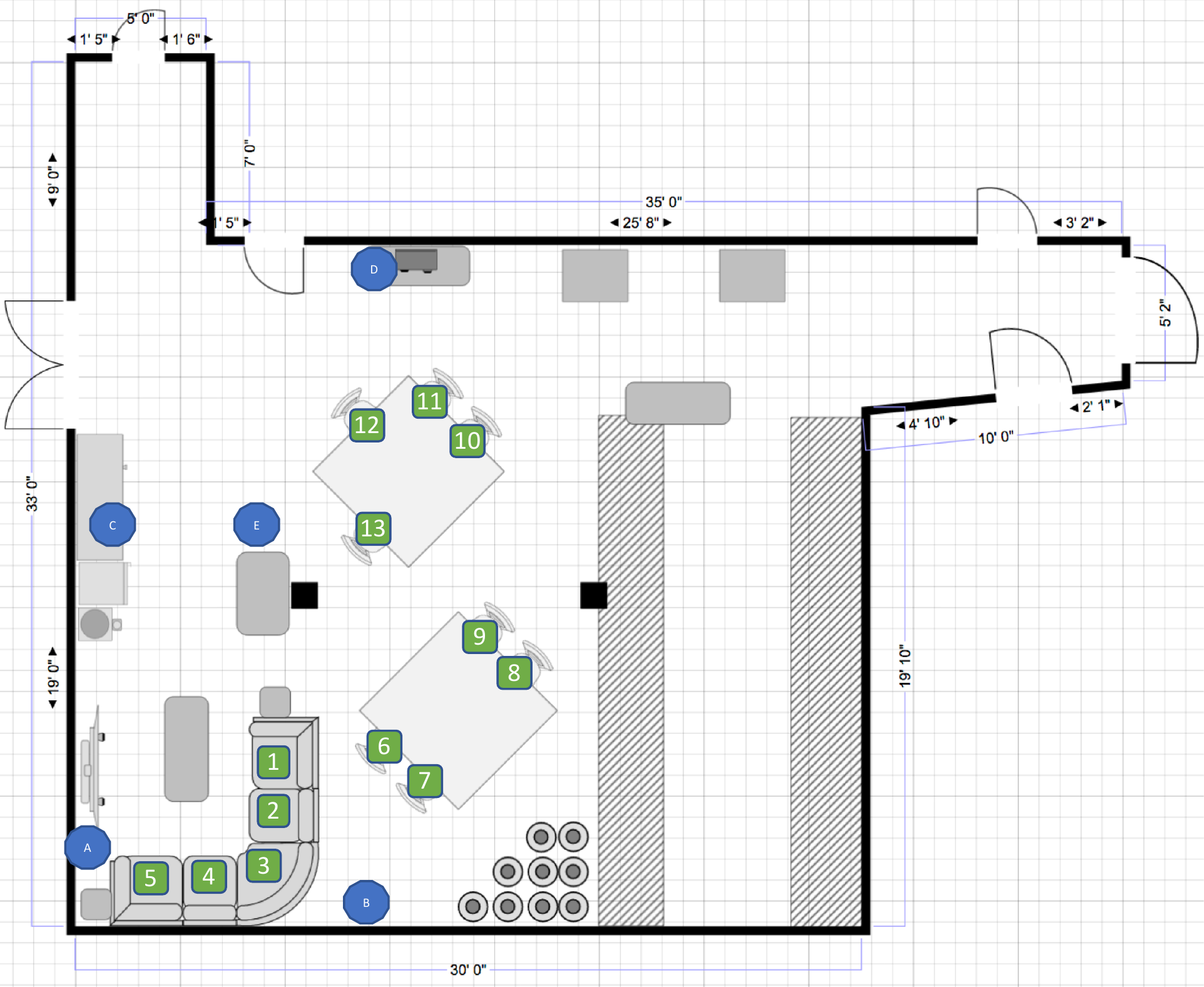}
\includegraphics[width = 3 in]{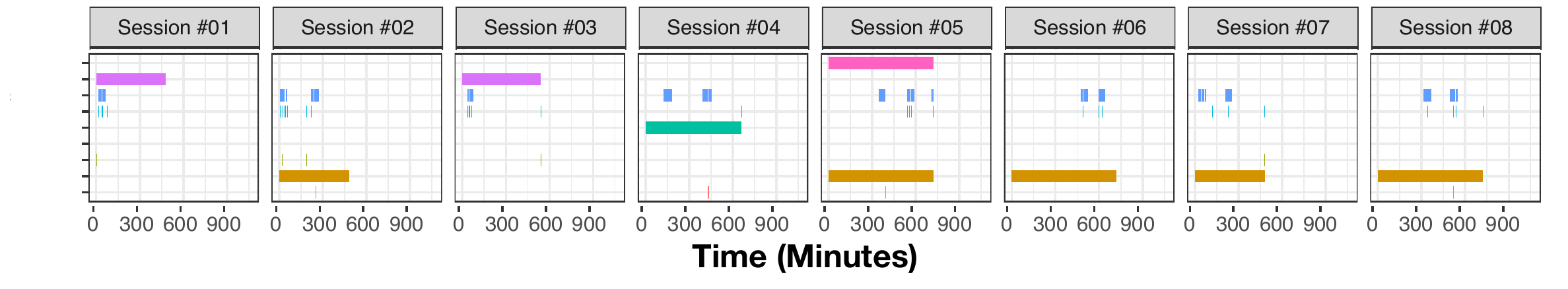}
\caption{Upper panel: Schematic diagram of Room  3,  which was one of two rooms used for evaluation.  The electret microphones are located at the blue circles, while the green squares indicate possible talker locations.  Lower panel: Timeline of live speech (blue) and distractors: traffic (pink), TV (magenta), radio (green), and recordings from LDC databases (orange).} 
\label{fig:room}
\end{figure}

\textbf{Organization of this paper.}  In the next section we describe \emph{STRFNet}, a system developed by CMU based on deep learning principles that was designed specifically to optimize performance on the 2019 JHU/APL VTD assessment. We then describe and discuss the performance of the system on VTD data provided by SRI.  To complement the VTD task, we also evaluate our system on spoofing detection using the Logical Access (LA) data in the recent ASVspoof 2019 challenge \cite{Todisco2019}.

\section{The STRFNet System}

\begin{figure}[ht]
\centering
\includegraphics[width = .5\columnwidth]{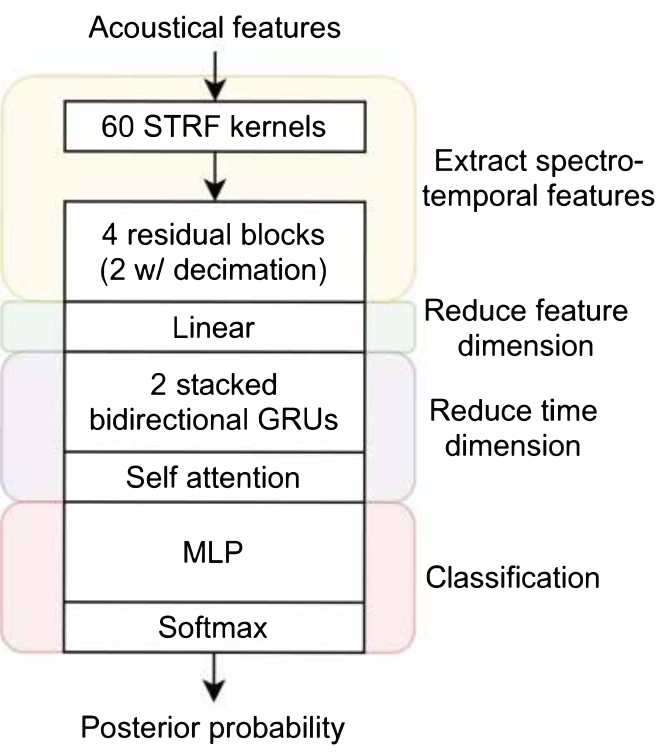}
\includegraphics[width = .4\columnwidth]{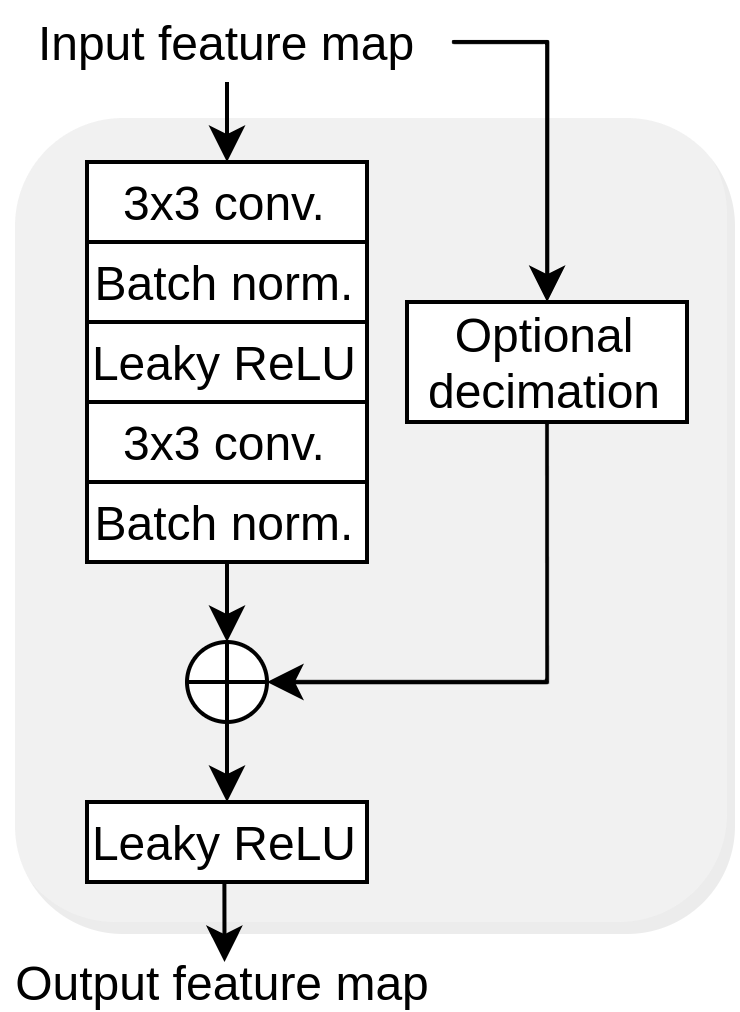}
\caption{Block diagram of the proposed system (left) and the zoomed-in view of one residual block (right).} 
\label{fig:SysBlox}
\vspace{-3 mm}
\end{figure}

Our design of the VTD system is motivated by three key assumptions: (1) the \ac{STRF}s can reliably extract speech-specific features in the modulation domain under adverse conditions, (2)  generic convolutional kernels are needed to extract patterns from the response to the \ac{STRF}, and (3) long temporal patterns might reveal suprasegmental features such as prosody that could potentially discriminate speaking styles in live speech from broadcast speech. Figure \ref{fig:SysBlox} describes the overall organization of the STRFNet system based on deep learning principles. After the extraction of acoustic features, the major components of the processing are: (1) a layer of 2-dimensional convolutional kernels where each is re-parameterized as an \ac{STRF}, (2) a series of convolutional layers with residual connections that further extract spectro-temporal features, (3) a stacked bidirectional \ac{GRU} with self-attention that compresses the time dimension, and (4) a series of feed-forward layers which end with a softmax normalization leading to a single probability of live speech. We describe all of these stages in greater detail in the paragraphs below.

\textbf{Initial signal processing.}
We considered three front ends for increased flexibility in initial processing: the \ac{CQT} \cite{Brown91,BrownPuckette92}, the log-mel spectrogram, and a bank of Gammatone filters \cite{PattersonEtAl95} implemented as convolutional kernels where each kernel has a learnable filter bandwidth and multiplicative gain. For the latter, the unit sample responses of the filters are all truncated at 513 samples to speed up training.  This option was motivated by the benefits gained by kernelizing filter responses \cite{LoweimiEtAl19,RavanelliBengio17} in a \ac{CNN}. While the choice of \ac{CQT} was motivated by the later \ac{STRF} processing in which the spectral modulation template assumes a fixed logarithmic frequency spacing, we found that the mel frequency spacing performed equally well on the tasks we considered. Moreover, we did not observe benefits from having learnable bandwidth and filter gains provided in the Gammatone filterbank. Therefore, we chose the computationally-efficient log-mel spectrogram.

\textbf{The learnable STRF layer and residual blocks.}
To  extract speech-specific spectro-temporal features in noisy and reverberant environments effectively, we proposed a convolutional layer in which each kernel is constrained to be an STRF \cite{ChiEtAl05} that models modulation on a log-frequency/linear-time scale. We used the same seed functions and the parameterization process proposed by Chi \etal \cite{ChiEtAl05} and defined the learnable upward and downward-drifting discrete STRFs as follows.
\vspace{-0.05in}
\begin{equation}\label{eqn:discrete-strf}
\begin{aligned}
    \resizebox{0.9\columnwidth}{!}{$\text{STRF}_{N,K}[n, k; \Omega, \omega, \phi, \theta]_{\Uparrow} = \text{STRF}(\frac{n}{N},\frac{k}{K}; \Omega, \omega, \phi, \theta)_{\Uparrow}$}\\
    \resizebox{0.9\columnwidth}{!}{$\text{STRF}_{N,K}[n, k; \Omega, \omega, \phi, \theta]_{\Downarrow} = \text{STRF}(\frac{n}{N},\frac{k}{K}; \Omega, \omega, \phi, \theta)_{\Downarrow}$}
\end{aligned}
\end{equation}
where the continuous \ac{STRF} functions correspond to the definition in \cite{ChiEtAl05}, $N$ is the frame rate in Hz, and $K$ is the number of frequency channels per octave. During neural network training, the two modulation frequencies ($\Omega$ and $\omega$) and two characteristic phases ($\phi$ and $\theta$), each referring to spectral and temporal processing, respectively,  were adapted by gradient backpropagation. This definition requires a finite time and frequency support, respectively; we discuss our own choices below.

In calculating the discrete STRF, the discrete Hilbert transform will be used in place of its continuous-time counterpart. Since the filter that produces the analytic function of an input signal has infinite impulse response, approximations need to be made so that the gradients backpropagate efficiently during training. For this reason, we design a finite impulse response Hilbert transform filter by the simple frequency-sampling method \cite[p.~394]{lim1987advanced}.
The M-point discrete Fourier transform  causes the analytic signal to be time-aliased to some extent; we use $M=512$ for all our experiments.

To enhance the extracted spectro-temporal features, we follow the \ac{STRF} convolutional layer with four residual blocks, whose architecture is illustrated in the right figure of Figure~\ref{fig:SysBlox}. This design is motivated by the success of residual connection in image recognition \cite{he2016deep} as well as its successful application in spoofing detection \cite{Cai2019,Alzantot2019DeepRN}.


%

\textbf{Bidirectional \ac{GRU}s with self attention.}
After spectro-temporal feature extraction through the \ac{CNN}s, features of each time frame are aggregated, flattened, and passed through a fully connected layer for further reduction of dimensionality.  Then, we use a stack of two \ac{GRU}s to learn long temporal patterns.  Following the GRUs, we use a self-attentive pooling layer \cite{Cai2018ExploringTE, India2019SelfMA} to compress the time dimension.  The self attention achieves this by using a trainable layer that assigns a learnable weight to each time frame and then performs a weighted average to obtain a single feature vector.  

After temporal modeling and the self-attentive pooling, the feature vector is passed through a one-hidden-layer \ac{MLP} with two dimensions for the output.  Finally, softmax is applied to obtain the posterior probability of a given segment being Live Speech. An implementation of the STRFNet system can be found at \url{http://www.cs.cmu.edu/~robust/code.html}. 

\section{Evaluation Tasks and Datasets}
\textbf{Datasets.}
We evaluated our proposed system on the VTD task and the spoofing detection tasks in the most recent ASVspoof 2019 challenge \cite{Todisco2019}. In this section, we describe the data we used as well as our evaluation plan.

As described in the introduction, SRI collected the data specifically for the VTD task.  The dataset consists of 34 recording sessions, 20 of which are used for training, 8 are used in development, and 6 are used for evaluation. The training data were recorded in Room 1 and Room 2 and the development data and evaluation data were recorded in Room 3 and Room 4, respectively.  Each recording session lasted an average of 9 hours,  live speech was only present for an average of 72 minutes scattered throughout the session.  The difficulty of this task comes from the fact that live speech was only present for a small fraction of each session and was often time-overlapping with  distractor audio such as television, radio broadcasts and non-broadcast-style speech from Linguistic Data Consortium (LDC) databases at low \ac{SNR}s.  The lower panel of Figure~\ref{fig:room} summarizes the live speech and distractor audio time regions for the development room along with a schematic diagram of the room.  Each of these sessions were recorded from 5 far-field microphones dispersed around the room.  During testing, we only have access to one of the microphones at a time.  
In total, the dataset comprised of roughly 800 hours, 400 hours, and 400 hours of audio recordings in the training data, development data, and evaluation data, respectively, and were all sampled at 11,025 Hz.

Although the STRFNet system was developed for the purpose of VTD, we wanted to test our system and evaluate the benefits of our proposed learnable STRF layer on a similar task that was publicly benchmarked.   We chose the ASVspoof 2019 challenge dataset \cite{wang2019asvspoof} (ASVspoof henceforth) since their tasks, both logical access (LA) and physical access (PA), are similar to VTD.  To better match the more challenging acoustical conditions of the VTD task we added VTD distractor audio to the ASVspoof data and downsampled the data   to 11,025 Hz. 


%

 We found in our pilot experiments that our baseline system performed comparably to  reported results obtained from the original data \cite{Todisco2019} on both the LA and PA tasks. However, the performance for the PA task became much worse  after the audio was downsampled and combined with VTD distractor audio. This suggests that the good benchmark performance on the PA task \cite{LiEtAl17} was enabled by subtle spectral distinctions that do  not survive decimation and additive noise. For these reasons  we  evaluate only on the LA task (ASVspoof-LA henceforth).  The modified ASVspoof-LA dataset consists of 24 hours, 24 hours, and 63 hours of training, development, and evaluation data, respectively, and 44 hours of VTD distractor audio that are randomly sampled and added.


\textbf{Evaluation metrics.}
For the VTD task, the systems were evaluated using the Detection Cost Function (DCF) proposed by the organizer as the primary metric and  Equal Error Rate (EER). The proposed DCF is a weighted average of the probability of false alarm ($P_{FA}$) and probability of a miss ($P_M$), with  misses weighted three times as much as false alarms.  Both probabilities are normalized according to the duration of segments that are labeled as Live Speech versus Distractor audio.  
For the ASVspoof-LA task, we evaluated the systems using the EER at various \ac{SNR}s.  When evaluating on clean speech only, we also used the tandem Detection Cost Function (t-DCF) \cite{Kinnunen2018tDCFAD}, the primary metric in the AVSspoof 2019 challenge, which combines the performance of both automatic speaker verification (ASV) and spoofing detection.  Since the provided ASV score was obtained on clean speech, we did not calculate the t-DCF when evaluating on noisy data.

\section{Experimental Procedures and Results}
In this section, we describe the experimental procedures for both tasks and discuss the results.

\textbf{Feature front end and data augmentation.}
As described earlier, we used the log-mel spectrogram as the  front end. The spectrogram was obtained using a 20-millisecond Hamming window with 50\% overlap between frames and a 512-point discrete Fourier transform. In all our experiments, 40 mel bands were used for the VTD task and 80 were used for the ASVspoof-LA task.  Logarithmic power compression using natural logarithm was applied subsequently.

To help our systems generalize to unseen speech and distractor audio, we applied the SpecAugment \cite{googlespecaug} method to the input feature during training.  SpecAugment is a simple data augmentation method that performs time warping and randomly masks entire frequency bands and time frames of the input spectrogram representation before passing it as input to the system.  SpecAugment was initially proposed to improve  automatic speech recognition \cite{googlespecaug}, but we found it to be helpful for both the VTD and ASVspoof tasks.  

\begin{table}
\centering
\caption{System configurations for VTD and ASVspoof tasks. Subscript $S$ denotes static \ac{STRF}s. The remainder of the systems are identical and follow Figure~\ref{fig:SysBlox}.}
\begin{tabular}{l|l|l|l}
System & First layer (VTD/ASVspoof) & \# Kernels \\
\hline
$\text{CNN}_K$ & 2DConv & $K$ \\
$\text{Hybrid}_S$ & -- / 2DConv+STRFConv & -- / 60+60 \\
Hybrid & -- / 2DConv+STRFConv & -- / 60+60 \\
$\text{STRFNet}_S$ & STRFConv / -- & 60 / -- \\
STRFNet & STRFConv / -- & 60 / --
\end{tabular}
\label{table:ASVspoof-noise-config}
\vspace{-4mm}
\end{table}

\textbf{Baseline systems.}
As shown in Table~\ref{table:ASVspoof-noise-config}, we included five types of baseline systems to study the effectiveness of the learnable \ac{STRF}s. In the first convolutional layer, $\text{CNN}_K$ consists of generic kernels (2DConv); Hybrid systems consist of both generic and \ac{STRF} kernels (STRFConv); STRFNet consists of only \ac{STRF} kernels. The conventional $\text{CNN}_K$ was tuned to produce competitive results compared to those reported in the official challenge \cite{Todisco2019}.  (The subscript $K$ refers to the number of kernels in the first layer in a given system.) 
After obtaining satisfactory baseline results on ASVspoof-LA, we matched the number of kernels for the other systems. The \ac{STRF}s in $\text{Hybrid}_S$ and $\text{STRFNet}_S$ were initialized with random parameters that were fixed during training. Specifically, the temporal and spectral modulation frequencies are uniformly sampled over $[0, 10)$ Hz and $[0, 0.2)$ cycles per channel, respectively, while characteristic phases are uniformly sampled over $[0, 2\pi)$. For the VTD task, the \ac{STRF} has a time support of .5 seconds and spans 15 frequency channels. For the ASVspoof-LA task, the \ac{STRF} has a time support of .25 seconds and spans 7 frequency channels.

\textbf{Training and evaluation procedure.}
To train our systems for the VTD task, we first randomly selected 5-second audio segments and computed a log-mel spectrogram. Next, we normalized each channel by subtracting the mean and dividing by the standard deviation of the channel across a single segment.  We then applied SpecAugment to the normalized features and used them as input to the systems. During testing, the systems took in one 5-second segment at a time and output raw posterior probability scores for that segment.   Before we calculated the DCF, we made a binary decision on each posterior probability score for each segment based on a fixed threshold. We then  applied a postprocessing step that re-labels as live speech selected brief segments that had been hypothesized as distractor audio.  We used the fixed threshold for each specific system that provided the lowest DCF for that system.  To calculate an EER that reflects the post-processing step used in the DCF procedure, we also apply the postprocessing step before the false positive rates and false negative rates are calculated.

To train our systems on the downsampled ASVspoof-LA data, we mixed the speech in every training file with a segment of randomly sampled VTD noise at global SNRs  of $5, 10, 15, 20, 25, 30, 40$ dB, selected randomly.    We only applied SpecAugment to the input features during training.  The VTD distractor audio for evaluation was disjoint from the training data.  We evaluated the EER of all systems at more finely  sampled \ac{SNR}s.  Both the EER and t-DCF were calculated when evaluating our systems on clean speech.

All systems were trained using the Adam optimizer \cite{adam} with a learning rate of $10^{-4}$ and a batch size of 64 in PyTorch \cite{pytorch}.  For both tasks, we stopped training when the performance on the development set stopped improving for multiple epochs.  Each system has about 2.4 million trainable parameters. 


\textbf{Baseline results.}
Table~\ref{table:Specaug-results} shows ASVspoof-LA results for our baseline system $\text{CNN}_{60}$. When trained with SpecAugment using the original data, $\text{CNN}_{60}$ produces a t-DCF  of .091, which is comparable to that of the $4^{th}$ place system in the official 2019 challenge for this task \cite{Todisco2019}.  We used the configuration highlighted in gray to train all our systems.

\begin{figure}[t]
\centering
  \includegraphics[width=2.55 in]{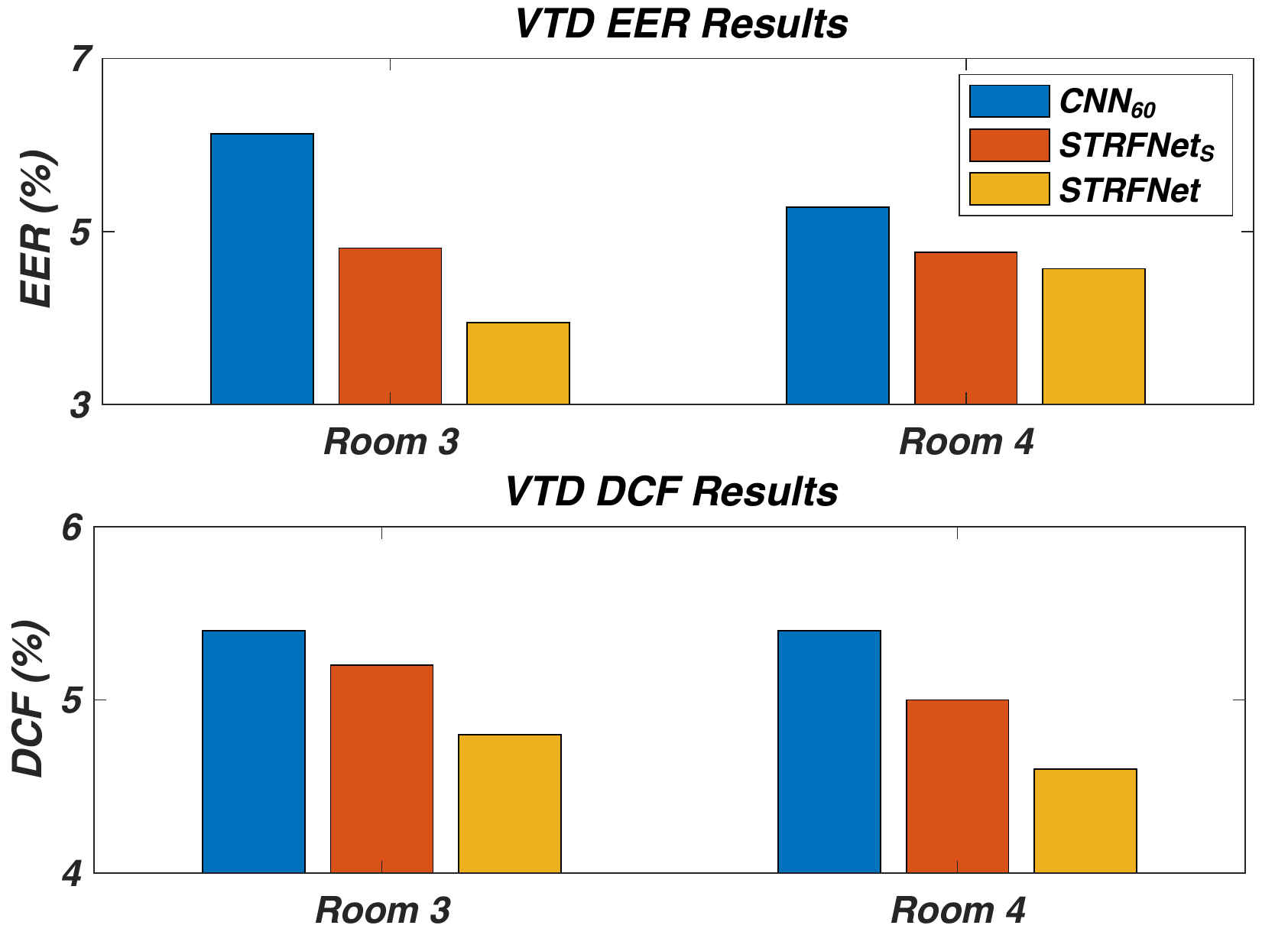}
  \caption{EER and DCF scores for the VTD task.}
  \label{fig:VTD_results}
  \vspace{-3 mm}
\end{figure}

\textbf{VTD results and discussion.}
 Figure~\ref{fig:VTD_results} summarizes the performance of the systems on the VTD task.  
 The proposed learnable STRFNet system outperforms both the baseline  $\text{CNN}_{60}$ system and the  $\text{STRFNet}_S$ system (which does not adapt) in both Rooms 3 and 4.  Our results show that replacing a generic convolutional layer with a static \ac{STRF} layer reduces both the DCF and EER on evaluation Room 4 by 5.7~\% relative and 9.9~\% relative, respectively.  Furthermore, we can obtain additional improvement by enabling the learnable component in the \ac{STRF} layers, further reducing the DCF and EER by 8.0~\% relative and 4.0~\% relative, respectively.
 
 We observed during the experiment that the STRFNet system was able to reject unseen noise with high confidence. However, the television and radio speech remains a challenging distractor for all systems, especially at low \ac{SNR}s. Selecting suitable \ac{STRF}s that discriminate different types of speech is an interesting area that we will explore in the future.

\begin{figure}[t]
\centering
  \includegraphics[width=2.85 in]{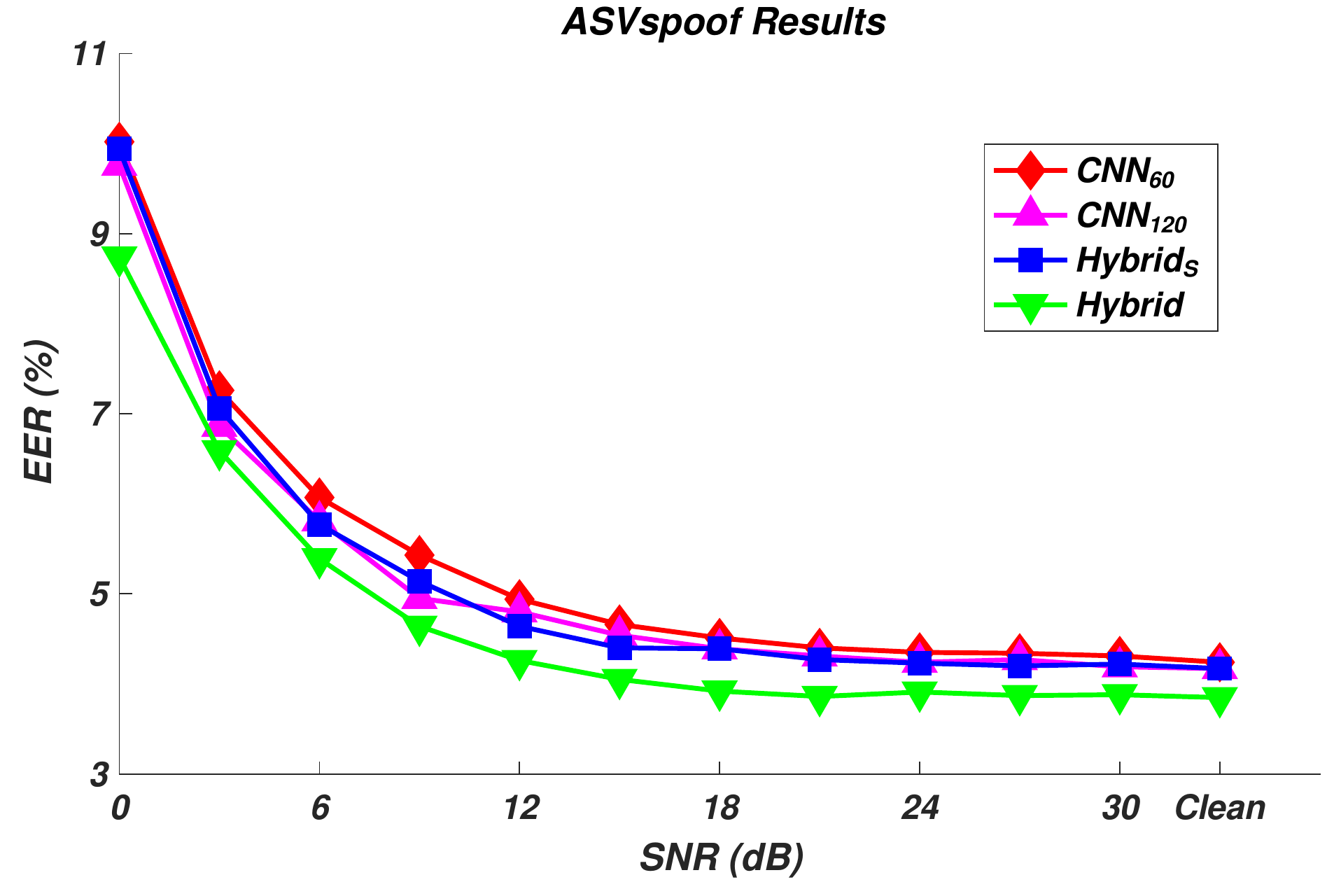}
  \caption{Equal error rates from ASVspoof-LA task.}
  \label{fig:EER_ASV}
\end{figure}

\textbf{AVSspoof results and discussion.}
The performance of each system on the downsampled ASVspoof-LA data with VTD distractor audio added at various \ac{SNR}s is shown in Figure~\ref{fig:EER_ASV}.  
Our  Hybrid system outperforms the original baseline $\text{CNN}_{60}$ by an average of 11.68\% relative EER and outperforms the two comparable baseline systems $\text{CNN}_{120}$ and  $\text{Hybrid}_S$  by an average of 8.61\% and 8.56\% relative EER, respectively. 
Our results show that our Hybrid model benefits from having both generic convolutional kernels and our proposed learnable STRF kernels in the first layer. During the experiments, we observed that STRFNet did not perform well on this task. Nevertheless,  the Hybrid system both performed the best and was the most robust to unseen noise and synthesis methods. This suggests that the \ac{STRF}s effectively reject distractor noise, but are by themselves not sufficient for discriminating real from synthetic speech.  

\begin{table}

\scalebox{1}{
\centering
\begin{tabular}{l|l|l|l|}
SpecAug. \cite{googlespecaug} & Samp. Rate (Hz) & EER (\%) & t-DCF  \\ 
\hline
Not Used            & 16000         & 6.49     & .116   \\
Used           & 16000         & 6.02     & .091   \\
Not Used            & 11025         & 8.20     & .187   \\
\rowcolor{lightgray} Used           & 11025         & 6.40     & .166  
\end{tabular}
}
\caption{Baseline performance of $\text{CNN}_{60}$ on ASVspoof-LA task.}
\label{table:Specaug-results}
\end{table}
In general, we find that degrading speech is an effective training technique that improves the systems generalization capabilities to the unseen spoofing techniques in the evaluation data. As shown in Table~\ref{table:Specaug-results}, SpecAugment drastically improved the system trained on the downsampled data. In addition, adding VTD distractor audio to the downsampled data during training further improved the EER of all the systems when evaluating on all conditions, including clean speech.






\section{Conclusions}
In this paper, we  incorporate learnable spectro-temporal receptive fields (STRFs) in a deep neural network for the emerging task of voice type discrimination (VTD). We show that systems using the proposed learnable \ac{STRF}s in the first layer consistently outperform a competitive baseline using generic kernels for the VTD task and for the logical access task in the ASVspoof 2019 challenge \cite{Todisco2019}. We also show that the learning component of the \ac{STRF} kernel is essential for both robust spoofing detection at a wide range of unseen noisy environments and the VTD task. In the future, we plan on comparing the Gabor-based implementation of the STRF kernels \cite{MEYER2011753} to the implementation used in this paper \cite{ChiEtAl05}.  We will also  evaluate the robustness of the learnable STRFs for other tasks such as robust speech recognition and robust speaker identification.

\section{Acknowledgement}

This work was funded by the Johns Hopkins University Applied Physics Laboratory.
\newpage
\bibliographystyle{IEEEtran}

\bibliography{sternbib}

\end{document}